\documentclass[12pt]{spieman}  
\usepackage[english]{babel}   
\usepackage{amsmath,amsfonts,amssymb}%% zusaetzliche Mathe-Symbole
\usepackage{textcomp,gensymb}
\usepackage{graphicx}
\usepackage{epstopdf}
\usepackage[separate-uncertainty=true,multi-part-units=single]{siunitx}
\usepackage{float}
\usepackage{subcaption}
\usepackage{setspace}
\usepackage{tocloft}

\usepackage{etoolbox}
\usepackage{lmodern}
\title{Towards 10\,cm\,s$^{\boldsymbol{-1}}$ radial velocity accuracy on the Sun using a Fourier transform spectrometer}
\author[a,*]{Michael Debus}
\author[a]{Sebastian Schäfer}
\author[a]{Ansgar Reiners}
\affil[a]{Institut f\"ur Astrophysik \& Geophysik, Georg-August-Universit\"at, Friedrich-Hund-Platz 1, G\"ottingen, Germany}
\renewcommand{\cftdotsep}{\cftnodots}
\cftpagenumbersoff{figure}
\cftpagenumbersoff{table} 

\graphicspath{{plots/}} % Directory in which figures are stored
\definecolor{darkred}{rgb}{0.5, 0.0, 0.0}
\definecolor{darkred}{rgb}{0, 0.0, 0.0}

\begin{document} 
\maketitle
%\tableofcontents

\begin{abstract}
The IAG solar observatory is producing high-fidelity, ultra-high-resolution spectra (R$>$\SI{500000}{}) of the spatially resolved surface of the Sun using a Fourier Transform spectrometer (FTS). The radial velocity (RV) calibration of these spectra is currently performed using absorption lines from Earth's atmosphere, limiting the precision and accuracy. To improve the frequency calibration precision and accuracy we plan to use a Fabry-Perot etalon (FP) setup that is an evolution of the CARMENES FP design and an iodine cell in combination. To create an accurate wavelength solution, the iodine cell is measured in parallel with the FP. The FP can then be used to transfer the accurate wavelength solution provided by the iodine via simultaneous calibration of solar observations. 
To verify the stability and precision of the FTS we perform parallel measurements of the FP and an iodine cell. The measurements show an intrinsic stability of the FTS of a level of \SI{1}{\meter\per\second} over 90 hours. The difference between the FP RVs and the iodine cell RVs show no significant trends during the same time span. The RMS of the RV difference between FP and iodine cell is \SI{10.7}{\centi\meter\per\second}, which can be largely attributed to the intrinsic RV precisions of the iodine cell and the FP (\SI{10.2}{\centi\meter\per\second} and \SI{1.0}{\centi\meter\per\second}, respectively). This shows that we can calibrate the FTS to a level of \SI{10}{\centi\meter\per\second}, competitive with current state-of-the-art precision RV instruments. Based on these results we argue that the spectrum of iodine can be used as an absolute reference to reach an RV accuracy of \SI{10}{\centi\meter\per\second}.
\end{abstract}


{\noindent \footnotesize\textbf{*} \linkable{mdebus@phys.uni-goettingen.de} }

%% for initial submission
%\begin{spacing}{2}   % use double spacing for rest of manuscript
%\linenumbers

%\input{sections/intro.tex}

\section{Introduction} \label{sec:intro} 

Over the last decades much progress was made to improve precision astronomical radial velocity (RV) measurements. 
The current state-of-the-art of instrumental RV precision is at the level of \SIrange{10}{30}{\centi\meter\per\second} achieved by instruments such as ESPRESSO, \mbox{EXPRES} and NEID.\cite{Pepe2021,Blackman2020,Halverson2016} 
In recent years however, it became apparent, that one of the limitations for improved precision lies not in the instruments themselves, but rather in overcoming the uncertainty introduced by intrinsic stellar RV jitter. \cite{Meunier2010,Lagrange2010a} 
Efforts to improve models of stellar RV jitter are made by observing the Sun with precision RV instruments.\cite{Haywood2016,Dumusque2015,Strassmeier2018,Lin2022} The IAG solar observatory \cite{Schaefer2020a} is somewhat unique as it employs a Fourier transform spectrometer (FTS) rather than an ech\'{e}lle spectrograph commonly used in precision radial velocity measurements. Additionally, it has the capability to observe not only the integrated solar disc, but also small, spatially resolved fractions of the solar disc.\cite{Schaefer2020a}

The FTS has two major differences compared with ech\'{e}lle spectrographs. The solar spectra are recorded at an extremely high resolution of \SI{0.024}{\per\centi\meter} (corresponding to a resolving power of about \SI{500000}{} or better). Furthermore, there are no multiple diffraction orders, i.e. the spectrum is recorded in 1D and a single factor suffices to calibrate the complete frequency axis. The largest drawback of the FTS, its comparably low sensitivity, is not limiting in the case of solar observations. %The most recent dataset of the IAG solar observatory encompasses spectra from 14 different positions on the Sun with the SNR ranging from 650 (center) to 420 (limb) covering a broad wavelength range from \SIrange{420}{800}{\nano\meter}. \cite{ellwarth2023} These data show significant discrepancies to results obtained from radiative transfer codes, thereby providing an impetus and opportunity to improve these models.\cite{reiners2023}
Up until now, the IAG solar data are calibrated using telluric lines. \cite{ellwarth2023} This calibration method however is limited to an accuracy on the order of \SI{15}{\meter\per\second}.\cite{balthasar1982terrestrial} To improve the measurement precision and accuracy, we introduced a modified source chamber into the FTS \cite{Schaefer2020}, which enables us to simultaneously feed the solar light (\SIrange{400}{800}{\nano\meter}) in parallel with a calibration light source in a separate wavelength region from \SIrange{800}{1000}{\nano\meter}, combined by a dichroic mirror.
Laser frequency combs (LFC) and Fabry-P\'{e}rot etalons (FP) are calibration light sources typically employed in precision RV instruments as they provide dense, stable calibration lines over a broad spectral range.
Calibrating the FTS with an LFC would reduce the SNR in the solar data due to its relatively high intensity noise. The intensity variations of the LFC are directly recorded in the interferogram and by virtue of the Fourier transform this noise is distributed equally among all frequencies.\cite{braultdavis} The LFC also requires too high of a resolution to record double-sided (symmetric) interferograms, which is necessary for improved phase correction of the solar spectra. Additionally, the LFC requires constant maintenance and an hour of startup time and is thus not a good solution for robotic observations. Therefore, we aim to use an FP illuminated by LEDs in the calibration wavelength channel in parallel to the solar measurements and perform separate calibration with an iodine cell in the science channel to improve the absolute accuracy of our measurements.

In this work we investigate the RV stability of the FTS by measuring an FP and an iodine cell illuminated by a tungsten halogen lamp in parallel. By comparing two independent calibration light sources we can confidently rule out that common RV drifts are caused by the calibration light sources. Furthermore, this enables us to gauge the level of calibration precision and accuracy we can achieve with the combination of FP and iodine cell. The FP has the advantage of higher SNR and does not contaminate the part of the solar spectrum we are interested in. However, FPs exhibit an overall drift and a chromatic RV drift \cite{Terrien2021,Schmidt2022,Kreider2022} over time. Additionally, the absolute peak positions of the FP are not known, so it does not provide accurate measurements. Nonetheless, an absolute calibration can be achieved by calibrating the FTS with the iodine before or after the solar measurements with the FP measured in parallel. Accurate measurements are required to measure the absolute convective blueshift of the sun, which are thus far limited in accuracy \cite{Reiners2016b} or bandwidth. \cite{LoehnerBoettcher2018}

This paper is organized as follows. In section \ref{sec:setup} we briefly discuss the experimental setup used in this work. Section \ref{sec:temp} investigates the temperature stability of the setup. The RV measurements and results of the FP are presented in section \ref{sec:RV}. In section \ref{sec:conc} we discuss the results and provide an outlook.

\section{Setup} \label{sec:setup}
The setup used in this work consists of three main parts. The two light sources under test are an LED illuminated FP and a tungsten halogen lamp illuminated iodine gas cell. An FTS (Bruker IFS125HR) is used to record all high resolution spectra. 
Additionally, we use  \mbox{Pt-100} sensors with a 4-wire configuration read out by an Omega \mbox{PT104-A} for temperature monitoring.

\subsection{Fabry-Perot etalon} \label{sec:fpsetup}
The FP setup is an evolution of the CARMENES FP design. \cite{Schaefer2012} High RV-stability is achieved by placing the FP in a temperature stabilized vessel under high vacuum. The FP in our setup has a free spectral range (FSR) of \SI{3.6}{\giga\hertz}, a finesse of $\mathcal{F}$=7 and a soft coating optimized for the wavelength range \SIrange{790}{1010}{\nano\meter} (built by SLS Optics Ltd.). It is illuminated by two Prizmatix LEDs (MIC-LED-850V \& MIC-LED-940Z) combined with a dichroic and fiber fed (Ceramoptec \SI{200}{\micro\meter} core diameter octagonal fiber). 
Our FP setup is optimzied for a less efficient instrument with a higher resolving power and a different wavelength range with respect to the CARMENES design.
In addition changes to the vacuum vessel design were made to improve the thermal stability of the FP.
The three main design changes of the vacuum vessel are that the optical bench is thermally decoupled from the unstabilized lid, guiding sheets have been placed in the double-wall to ensure an even flow and distribution of the thermofluid and there is a heat shield between the unstabilized lid and the optical bench. A more detailed description of the setup can be found in Debus et. al. (2022)\cite{debus2022near}.

\begin{figure}[b]\centering
 \includegraphics[width=0.8\textwidth]{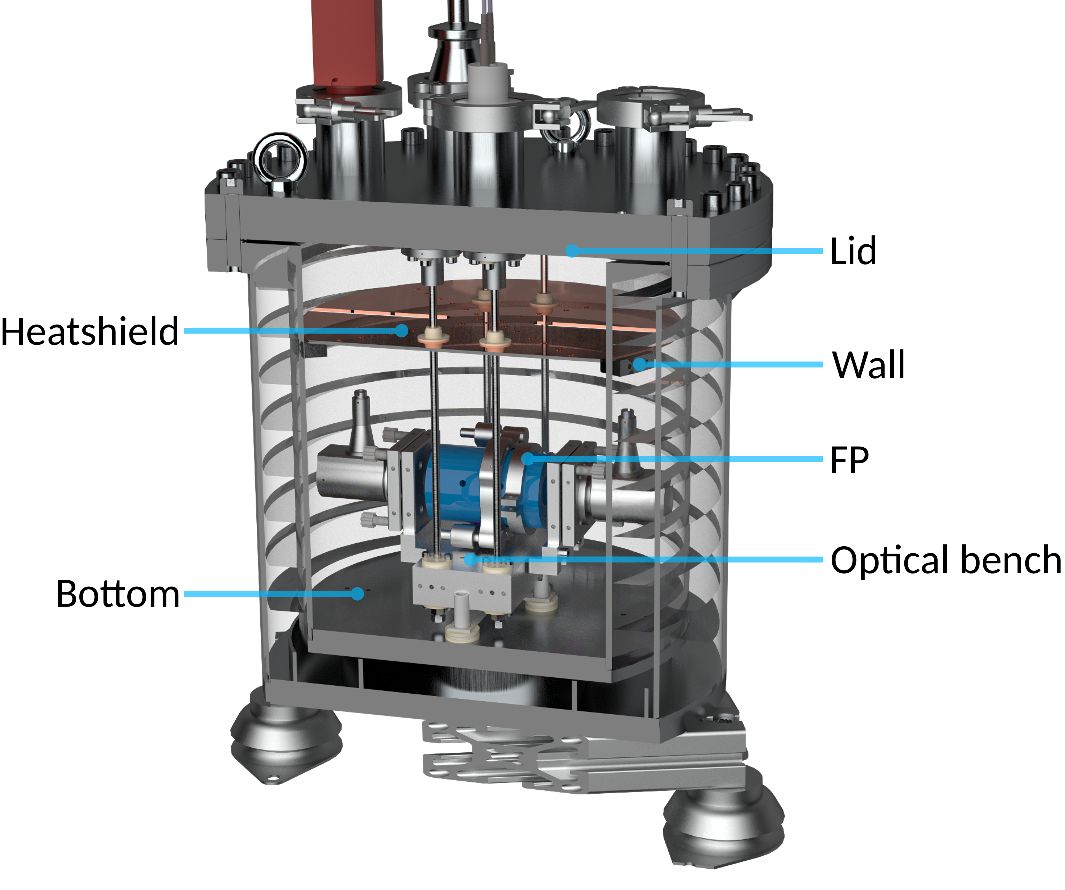}
 \caption{CAD-model of the vacuum vessel housing the optical bench of the FP. The Pt-100 sensor positions and their labels are indicated by the blue dots. The new machine feet for the vessel can also be seen.}\label{fig:cad}
\end{figure}

In addition to the aforementioned changes, the setup has been improved further since the description in Debus et al. (2022). First, the vacuum vessel has been placed in a Styrodur$^\text{\textregistered}$  enclosure for better insulation from changes in the room temperature (insulation thickness \SI{8}{cm}). Second, six Pt-100 temperature sensors have been placed at different locations inside the vacuum vessel (see Fig.\ \ref{fig:cad}). There are also \mbox{Pt-100} temperature sensors for monitoring the temperature inside the Styrodur$^\text{\textregistered}$ enclosure and the room temperature. 
Third, we exchanged the 3D-printed feet with commercially available feet for heavy machines (Ganter Norm GN 148) in order to improve the mechanical stability of the vacuum vessel. While these will in principle dampen vibrations from neighbouring setups this is not needed in our case.

\subsection{Iodine gas cell}
The iodine gas cell setup is quite simple% (see Fig.\ \ref{fig:iod}
, consisting of a commercial gas cell (Thorlabs \mbox{GC19100-I}) with a heater assembly (Thorlabs \mbox{GCH25-75}), wrapped in aluminum foil. A tungsten halogen lamp (OceanOptics HL-2000-HP-FHSA) is used for illumination. Light is coupled in and out with reflective collimators (Thorlabs \mbox{RC08SMA-P01} and \mbox{RC08FC-P01}) and $\varnothing$\SI{400}{\micro\meter} multi-mode fibers. For temperature monitoring a \mbox{Pt-100} sensor is placed inside the aluminum foil wrapping near the center of the gas cell.

% \begin{figure}[H]\centering
%  \includegraphics[width=0.8\textwidth]{
% iod_setup.jpg}
%  \caption{Photograph of the iodine cell setup. [xxx is this necessary?]}\label{fig:iod}
% \end{figure}

\subsection{Fourier transform spectrometer}

\begin{figure}[b]
\centering
 \includegraphics[width=\textwidth]{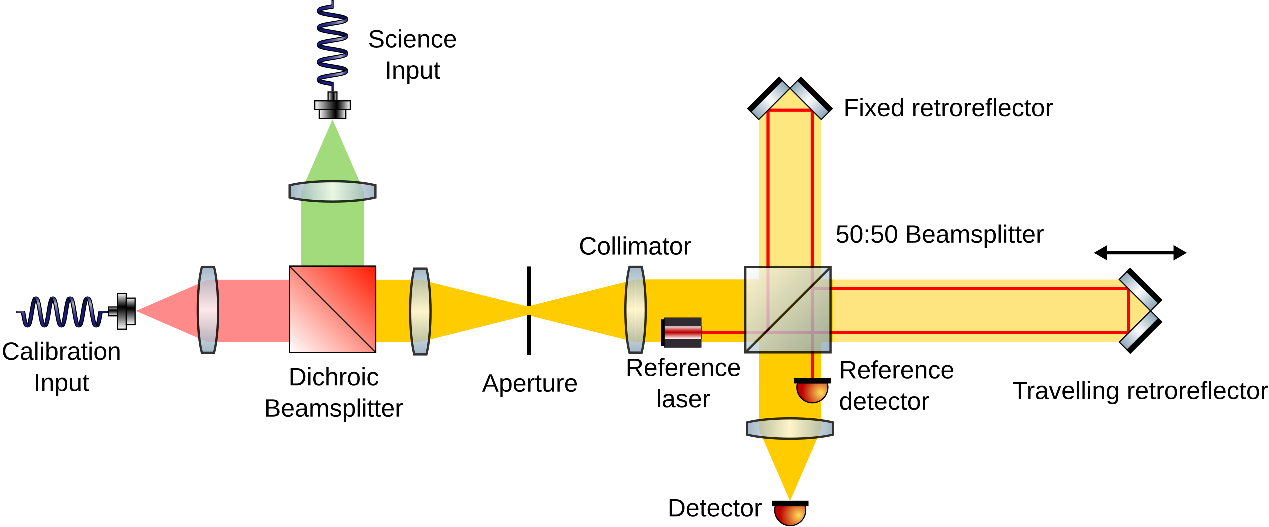}
 \caption{Schematic of the Fourier transform spectrometer setup. The modified source chamber can be seen to the left of the aperture.}\label{fig:fts}
\end{figure}

The FTS is built in the standard Michelson-interferometer configuration with a moving end mirror in one interferometer arm (see Fig.\ \ref{fig:fts}). %[xxx improve text here]
The measurement of intensity as a function of optical path length difference - called interferogram - is recorded at equal intervals corresponding to a quarter of the HeNe reference laser wavelength. The spectrum is obtained by calculating the Fourier transform of the interferogram. 

Possible error sources include a - potentially variable - misalignment between the science light and the HeNe-reference laser or a drift of the frequency of the HeNe-reference laser. These both lead to a stretching or squeezing of the interferogram, which can be described by a corresponding stretching or squeezing of the frequency axis of the spectrum by a single factor $k$:
\begin{equation}
 f_\text{true}=f_\text{measured}(1+k)
\end{equation}
Since the frequency axis only has one degree of freedom, we can calibrate it by using a calibration light source in a different spectral region than the light source of interest contrary to ech\'{e}lle spectrographs normally used in extreme precision RV measurements. 

\begin{figure}[htb]
\centering
 \includegraphics[width=\textwidth]{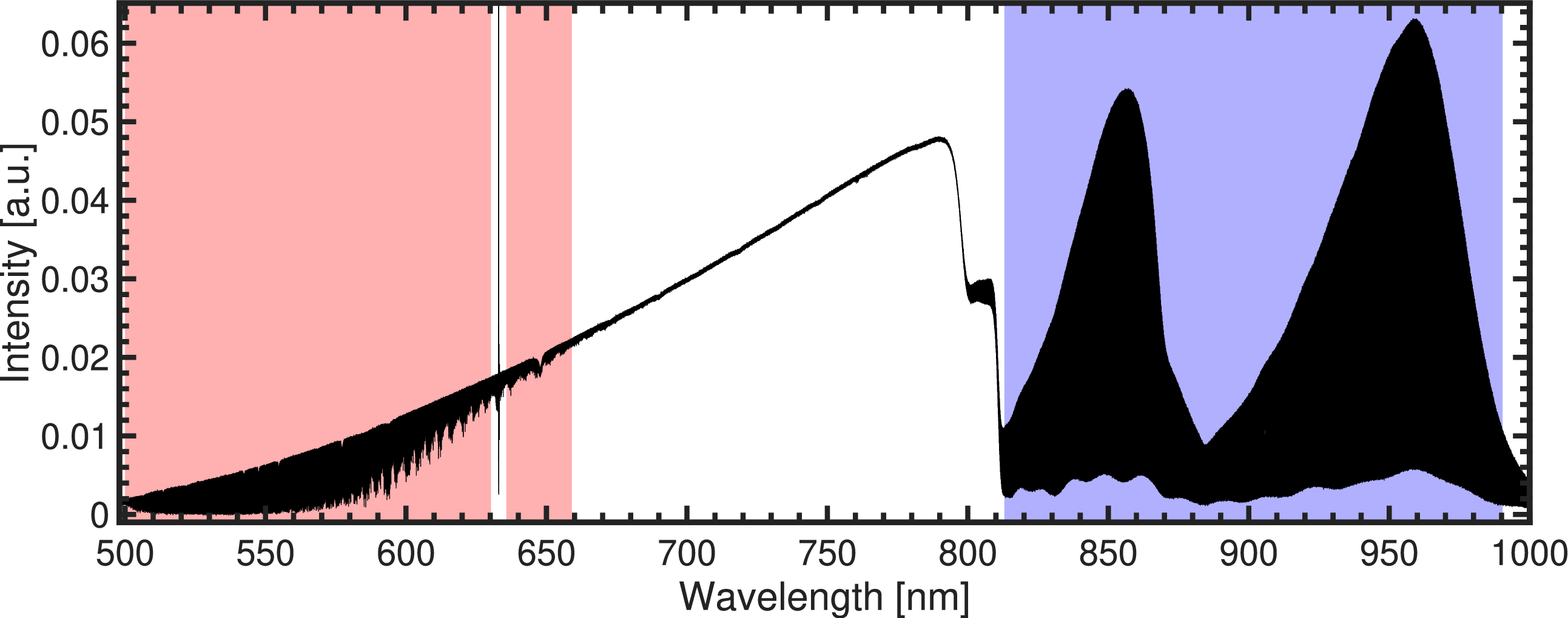}
 \caption{Example spectrum from the measurement run. The spectral regions used for the RV measurements are shaded in red for the iodine cell and in blue for the FP.}\label{fig:spec}
\end{figure}

Our FTS has a modified source chamber (for more detail see Schäfer et. al. \cite{Schaefer2020}) with two separate input fibers, the light from which can be combined with a dichroic beamsplitter allowing for a low wavelength channel ($<$ \SI{800}{\nano\meter}, science channel)  and a high wavelength channel ($>$ \SI{800}{\nano\meter}, calibration channel) to be measured in parallel (see left part of Fig.\ \ref{fig:fts}). The basic observing mode is to couple the integrated or resolved sun into the science channel and the FP into the calibration channel in parallel to compensate drifts of the FTS. In order to calibrate the absolute offset of the solar measurements we take measurements of the iodine cell in the science channel with the FP in the calibration channel in parallel. Figure \ref{fig:spec} shows an example spectrum of the combined light of the FP and the iodine cell measured with the FTS. In this work we investigate whether these two channels do drift against each other, which would prohibit tracking the science channel drift with the FP in the calibration channel. If the two calibration light sources do not drift against each other, the FP can be used to preserve the absolute accuracy of the iodine cell in the science channel for solar measurements.

\section{Temperature stability} \label{sec:temp}
The most critical part for achieving high precision RV calibration with an FP typically is the temperature stability of the setup. The main impact of a temperature change is the expansion of the Zerodur spacer. In our case this translates into a required temperature stability of  \SI{17}{\milli\kelvin} to reach a precision better than \SI{10}{\centi\meter\per\second}. While our FP setup is based on the established CARMENES design, it has significant improvements (see Sec.\ \ref{sec:fpsetup}). Additionally, the temperature stability of the CARMENES FP has thus far not been measured. We therefore investigate the thermal stability of our setup in the following.
 
\subsection{Vacuum vessel}
\begin{figure}[htb]
\centering
     \begin{subfigure}[c]{0.45\textwidth}
        \centering 
        \includegraphics[width=\textwidth]{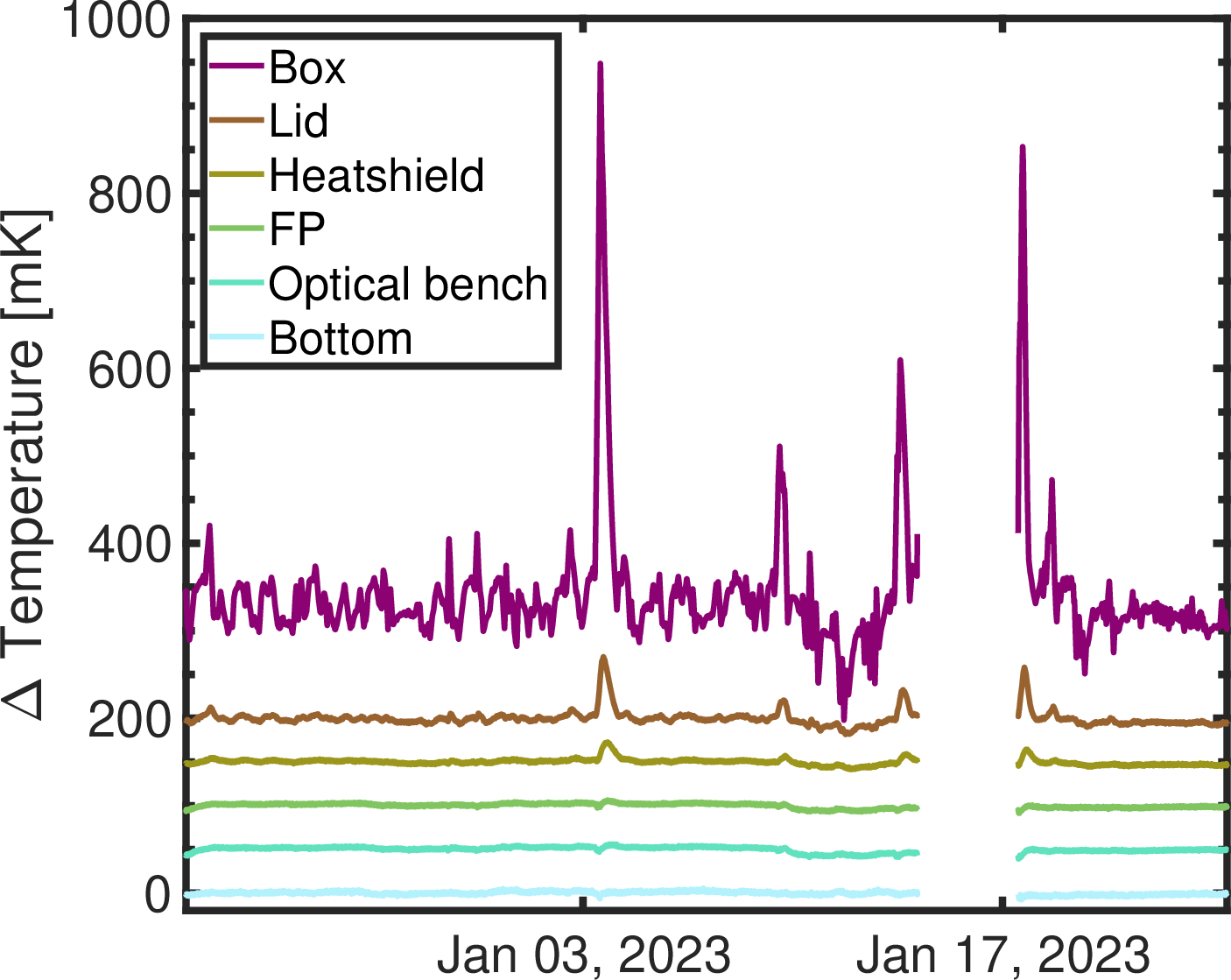}
        \caption{Ten minute moving mean of the temperature variations over a time span of 36 days. Curves are offset for visibility. } \label{fig:longtermtemp}
     \end{subfigure} \hspace{0.5cm}
     \begin{subfigure}[c]{0.45\textwidth}
        \centering \vspace{0.5cm}
        \includegraphics[width=\textwidth]{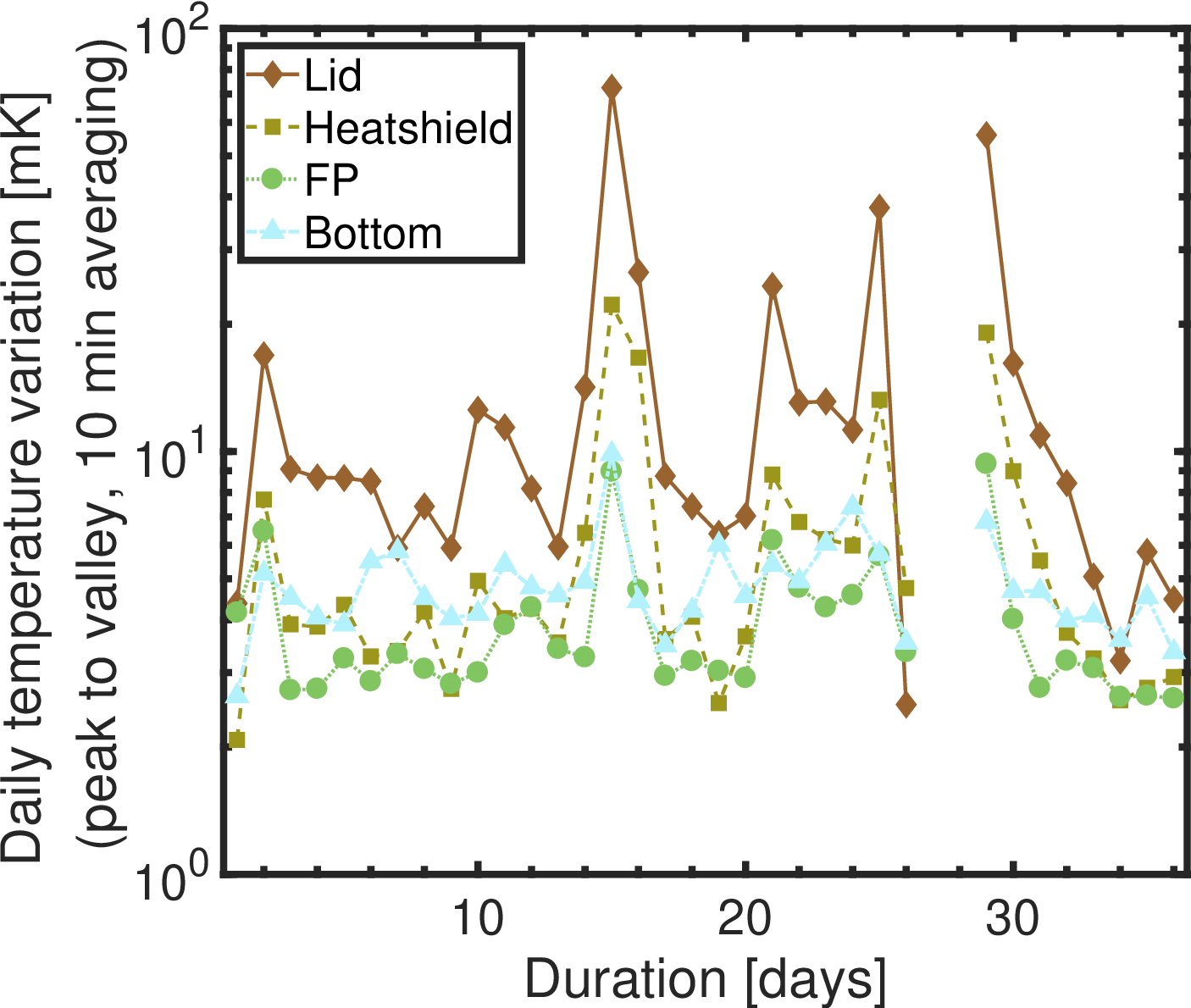}
        \caption{Maximum daily difference in the measured temperatures (ten minute moving mean) during the measurement period.}\label{fig:longtermdaylyvar}
     \end{subfigure}\vspace{0.3cm}
    \caption{ The ''Box''-sensor is located inside the Styrodur$^\text{\textregistered}$ enclosure but outside the vacuum vessel. The positions of the other Pt-100 sensors are shown in Fig.\ \ref{fig:cad}. The sensor for the vessel wall was omitted for better visibility as it shows a similar behavior as the vessel bottom.}
\end{figure}

We are monitoring the temperature in the FP vacuum vessel as well as the Styrodur$^\text{\textregistered}$ enclosure box and the laboratory itself. In a span of 36 days (with a two day gap due to a software crash) we see a peak-to-valley difference of \SI{15.1}{\milli\kelvin} in the 10 minute moving mean of the temperature of the sensor located on the FP (see Fig.\ \ref{fig:longtermtemp}). In the same period, the peak-to-valley difference in the room temperature was \SI{3.7}{\kelvin} while the peak-to-valley difference of the Styrodur$^\text{\textregistered}$ enclosure box temperature was \SI{751}{\milli\kelvin}. 

% \begin{figure}[htb]
% \centering
%  \includegraphics[width=\textwidth]{longterm_movmean.eps}
%  \caption{Ten minute moving mean of the temperature variations over a timespan of 34 days for the 6 sensors located inside the tank and the sensor in the Styrodur$^\text{\textregistered}$ enclosure box. Curves are offset for visibility. The positions of the Pt-100 sensors are shown in Fig.\ \ref{fig:cad}.} \label{fig:longtermtemp}
% \end{figure}

\begin{figure}[htb]
\centering
 \includegraphics[width=\textwidth]{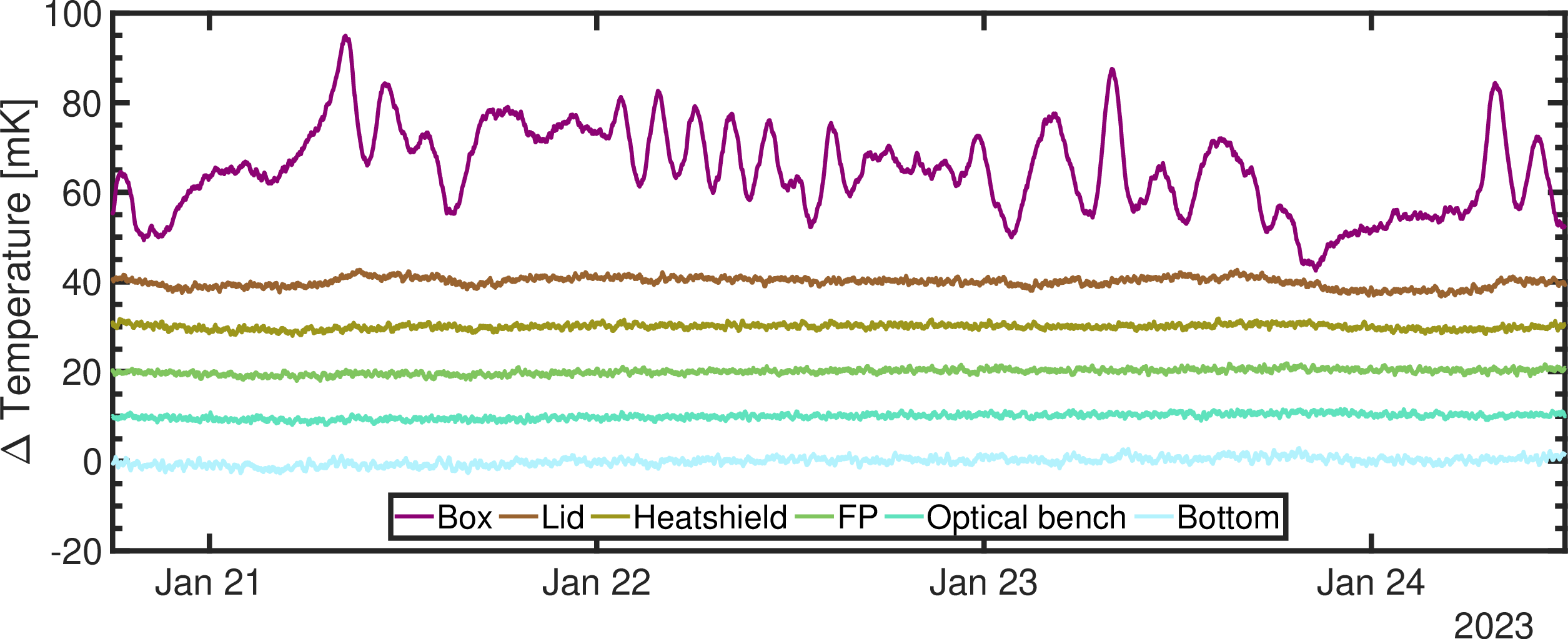}
 \caption{This zoom in of Fig.\ \ref{fig:longtermtemp} shows the ten minute moving mean of the temperature variations during the RV measurement period for the 6 sensors located inside the tank and the sensor in the Styrodur$^\text{\textregistered}$ enclosure box. Curves are offset for visibility. The positions of the Pt-100 sensors are shown in Fig.\ \ref{fig:cad}.} \label{fig:temp}
\end{figure}

Overall, the RV stability measurement period (see Sec.\ \ref{sec:RV}) was particularly calm  in terms of room temperature with an overall peak-to-valley difference of \SI{742}{\milli\kelvin}, leading to very stable temperatures in the vacuum vessel as well (see Fig.\ \ref{fig:temp}).
During these measurements, the peak-to-valley difference in the temperature measured at the FP (again in the 10 minute moving mean) was \SI{3.9}{\milli\kelvin}. This corresponds to the average daily peak-to-valley difference of the FP temperature in the long term of \SI{3.9}{\milli\kelvin} (see Fig.\ \ref{fig:longtermdaylyvar}). From the thermal expansion coefficient of Zerodur, the spacer material of the FP, we can expect a temperature induced RV drift of \SI{0.6}{\centi\meter\per\second\per\milli\kelvin} peak-to-valley for the FP and thus a maximum overall variation of \SI{2.4}{\centi\meter\per\second}. 

After a change in the vacuum vessel temperature it took about 24 hours for the temperature at the FP sensor to stabilize. The measured FP RVs however only stabilized after more than 8 days, indicating that the timescales at which the actual etalon thermalizes with respect to the sensor located at its aluminum casing is very long. This indicates, that the maximum temperature induced RV variation is even lower than the \SI{2.4}{\centi\meter\per\second} calculated from the sensor reading at the FP.

% \begin{figure}[htb]
% \centering
%  \includegraphics[width=\textwidth]{longterm_daylyvar10min_zoom.eps}
%  \caption{Maximum dayly difference in the measured temperatures (ten minute moving mean) during the measurement period for the 6 sensors located inside the tank and the sensor in the Styrodur$^\text{\textregistered}$ enclosure box. The positions of the Pt-100 sensors are shown in Fig.\ \ref{fig:cad}. The sensors for the optical bench and the vessel wall were omitted for better visibility as they show similiar behavior as the FP and vessel bottom, respectively.}\label{fig:longtermdaylyvar}
% \end{figure}

\subsection{Room temperature spikes}
\begin{figure}[htb]
\centering
    \begin{subfigure}[c]{0.45\textwidth}
        \centering \includegraphics[width=\textwidth]{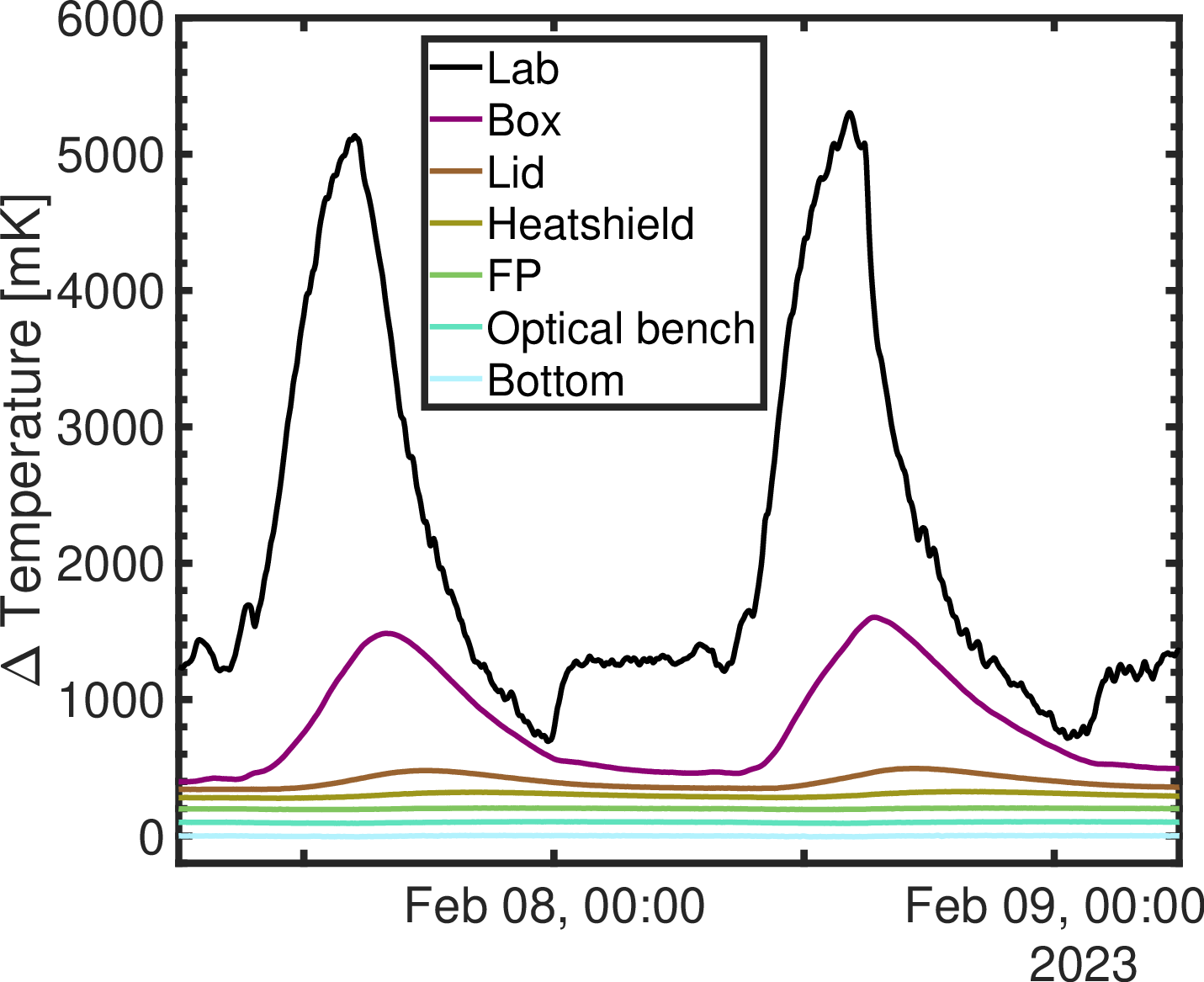}
         \caption{Temperature spike events. \label{fig:tempspikes}}
    \end{subfigure} \hspace{0.5cm}
    \begin{subfigure}[c]{0.45\textwidth}
        \centering \includegraphics[width=\textwidth]{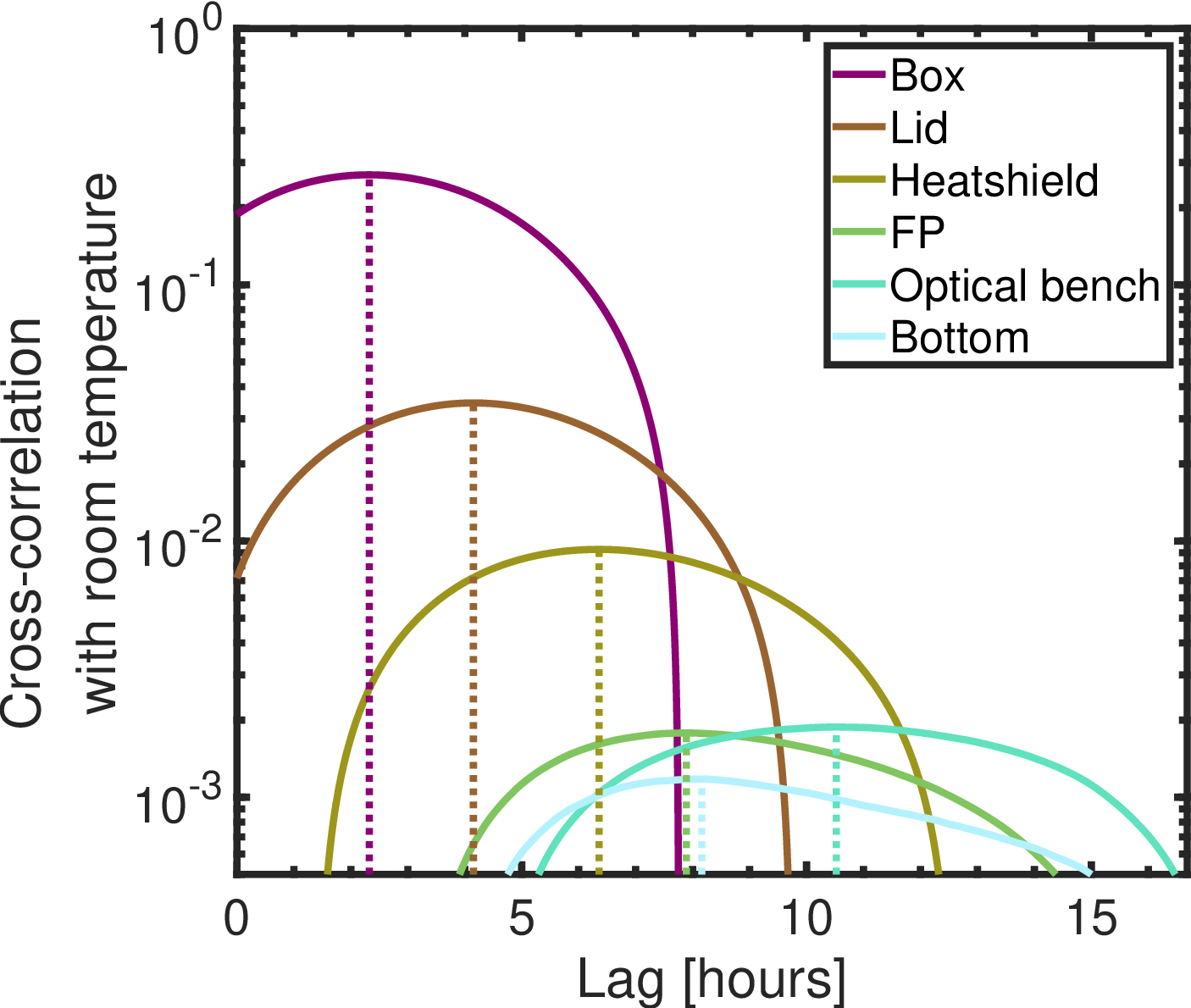}
        \caption{Cross correlation with room temperature.\label{fig:temp_xcorr}}
        %% lagtimes: 04:09:07   08:05:49   06:27:21   08:23:26   10:49:17   01:02:10   02:16:50        
     \end{subfigure}\vspace{0.3cm}
    \caption{Two events with an increase in laboratory temperature by \SI{4}{\kelvin} and the response of the overall system. Dotted lines indicate the lag at the maximum of the cross correlation. All cross correlation functions are normalized to the maximum of the autocorrelation of the room temperature. The positions of the Pt-100 sensors are shown in Fig.\ \ref{fig:cad}. }        
\end{figure}

We also have temperature data on two events during which the laboratory temperature rose by more than \SI{4}{\kelvin} on two consecutive days (see Fig.\ \ref{fig:tempspikes}). These data are instructive to show the thermal response of the different parts of the setup.

From the temperature spike dataset we can estimate the lag timescales as the temperature moves through the setup as well as the magnitude of the temperature dampening at each sensor position by cross-correlating the data from each sensor with the room temperature. Since the shapes of the temperature curves are not the same these are however only rough estimates. The results are shown in Fig.\ \ref{fig:temp_xcorr}. Unsurprisingly, the cross-correlation signal is strongest and has the lowest lag for the sensor inside the Styrodur$^\text{\textregistered}$ enclosure box at a little over 2 hours. The estimated temperature dampening factor and lag time at the FP-sensor are 562 and 8 hours. The optical bench with its high thermal mass takes even longer at 11 hours and a has a similar dampening factor of 532. Considering the dampening factor at the FP and the thermal RV sensitivity of the FP, this means that a change in room temperature of about \SI{11}{\kelvin} would be required to cause a \SI{10}{\centi\meter\per\second} shift in the FP-RVs.

% \begin{figure}[H]
% \centering
%  \includegraphics[width=\textwidth]{labtemp_xcorr.eps}
%  \caption{Cross correlation function of each temperature sensors data with the sensor recording the room temperature. Results are normalized to the maximum for better visibility, the corresponding normalization factor is shown in the legend.}\label{fig:temp_xcorr}
%  lagtimes: 04:09:07   08:05:49   06:27:21   08:23:26   10:49:17   01:02:10   02:16:50
% \end{figure}

\subsection{Iodine cell}

The iodine cell temperature shows an average daily peak-to-valley difference of \SI{163}{\milli\kelvin} and an overall peak-to-valley difference of \SI{402}{\milli\kelvin} over a timescale of five days. Comparing this to the expected temperature-induced RV drift of \SI{11}{\centi\meter\per\second\per\kelvin},\cite{Perdelwitz2018} we can expect a maximum daily amplitude of \SI{1.8}{\centi\meter\per\second} and a maximum overall amplitude of \SI{4.4}{\centi\meter\per\second} for the iodine cell RV stability. As it seems highly unlikely the iodine temperature has a measurable impact on the RV measurements, plots of the iodine temperature are omitted for the sake of brevity.

% 
% \begin{figure}[htb]
% \centering
%  \includegraphics[width=\textwidth]{RV_vs_Temp_spikes.eps}
%  \caption{Ten minute moving mean of the temperature variations during the temperature spikes in the laboratory. The RV variations of the iodine cell are also shown where available. The positions of the Pt-100 sensors are shown in Fig.\ \ref{fig:cad}.} \label{fig:tempspikes}
% \end{figure}

\section{Radial velocity stability}\label{sec:RV}
For the RV stability measurements we recorded 227 interferograms with an individual measurement time of 24 minutes, amounting to a total of more than 90 hours of measurement time. Light from the iodine cell was fed into the science channel, while the FP was fed into the calibration channel. Each interferogram is an average of ten measurements at a resolution of \SI{0.02}{\per\centi\meter} (corresponding to a resolving power of 500,000 at \SI{1000}{\nano\meter} and 1,000,000 at \SI{500}{\nano\meter}). All interferograms are double-sided, meaning that an equal optical pathlength difference (OPD) was scanned around the central interference peak at zero OPD. This simplifies and improves the phase correction as no approximation of the calculated phase is required. \cite{Griffiths2007} The FTS was evacuated to a level of \SI{0.19}{\milli\bar} throughout the measurements.

The radial velocities and their error estimates from each spectral region were calculated using SERVAL.\cite{Zechmeister2017} SERVAL determines the RV-shift with a least-squares fit of a Doppler-shifted template spectrum. The template is created in a first iteration by taking the highest SNR spectrum as an initial template. Each spectrum is shifted into the rest frame of that high SNR spectrum and then coadded to create the template.

\begin{figure}[tb]
\centering
 \includegraphics[width=\textwidth]{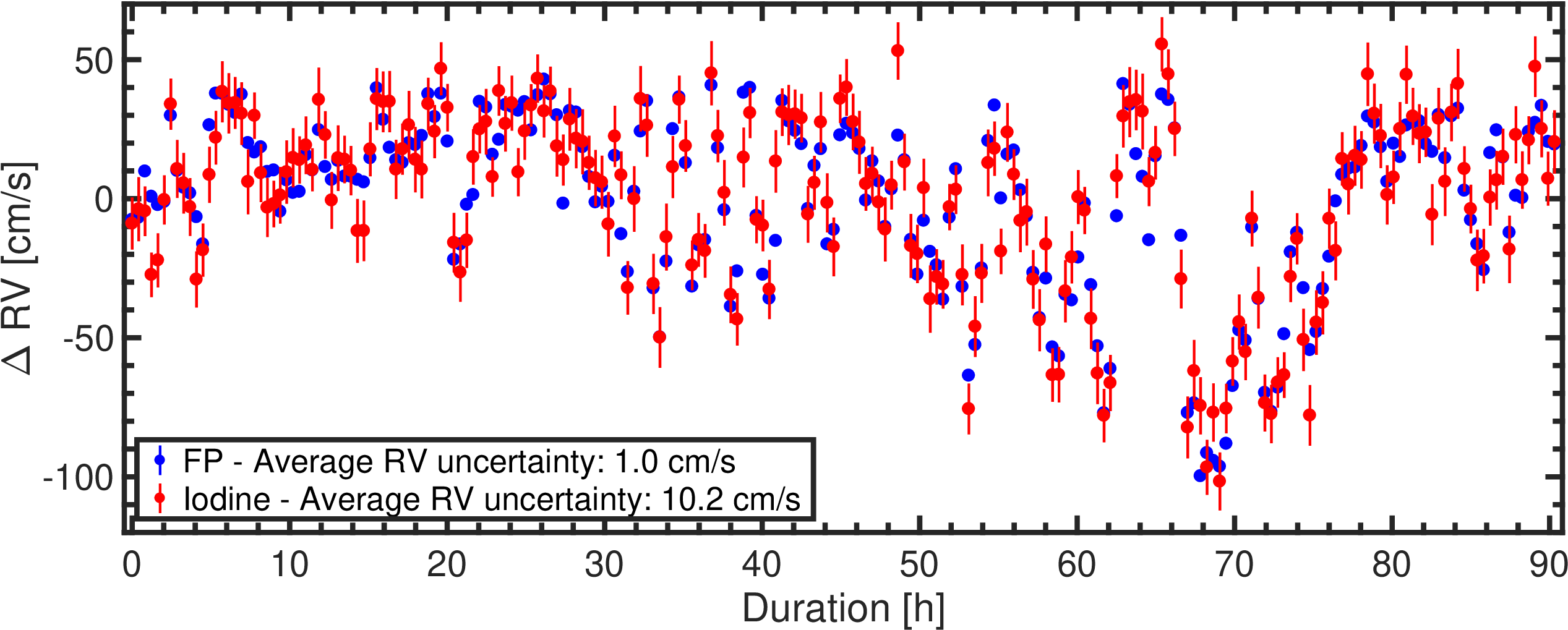}
 \caption{Measured radial velocity shifts for the FP and the iodine cell. Note that error bars for the FP measurements are at the \SI{1}{\centi\meter\per\second} level and thus not visible at this scale.}\label{fig:rvs}
\end{figure}

For the iodine we used the spectral region from \SIrange{498}{659}{\nano\meter}, omitting the vicinity ($\pm\,$\SI{3}{\nano\meter}) of the HeNe reference laser artifact around \SI{633}{\nano\meter} (see Fig.\ \ref{fig:spec}). For the FP we used the spectral region from \SIrange{813}{990}{\nano\meter}. Fig.\ \ref{fig:rvs} shows the relative RV-shift for each light source throughout the measurements. The average RV uncertainty in a single spectrum is \SI{10.2}{\centi\meter\per\second} for the iodine cell and \SI{1.0}{\centi\meter\per\second} for the FP. 

During the 90 hour measurement period, the measured RVs for both the FP and the iodine cell varied by less than $\pm\,$\SI{1}{\meter\per\second}. This is better than the specified HeNe reference laser stability of $\pm\,$\SI{3}{\meter\per\second} on a timescale of one hour (\SI{6}{\meter\per\second} on a timescale of one day). Most of the apparent drift is likely explained by changes in the HeNe reference laser frequency. Since our measurements took place in a relatively stable thermal environment the performance was probably better than the specification during our measurement.

\begin{figure}[tb]
\centering
 \includegraphics[width=\textwidth]{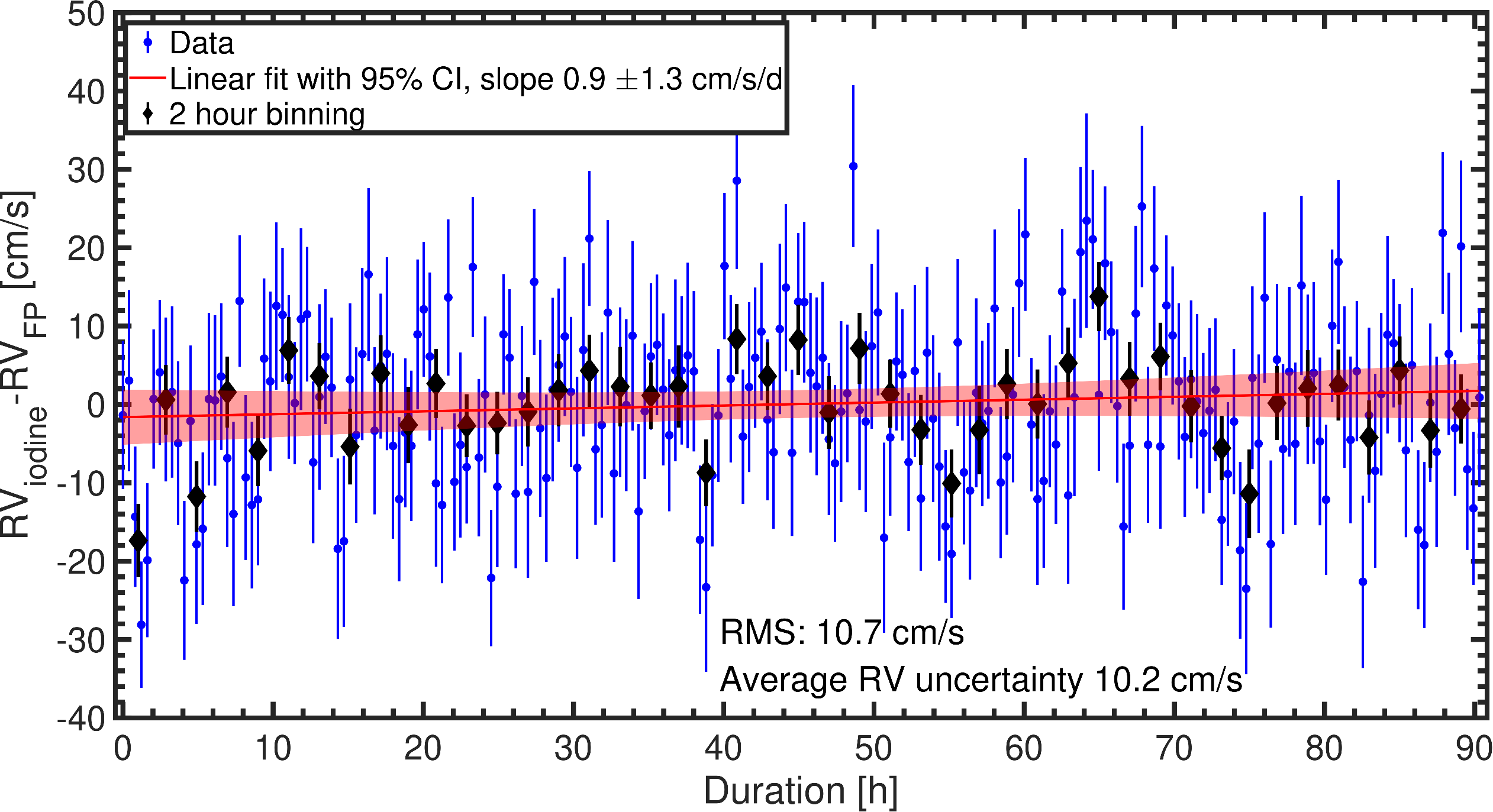}
 \caption{Difference between the measured radial velocities of the FP and the iodine cell. A linear fit to the data yields a non-significant slope. A 2 hour binning of the data also indicates no significant long term trends.}\label{fig:rvdiff}
\end{figure}

\begin{figure}[tb]
\centering
 \includegraphics[width=\textwidth]{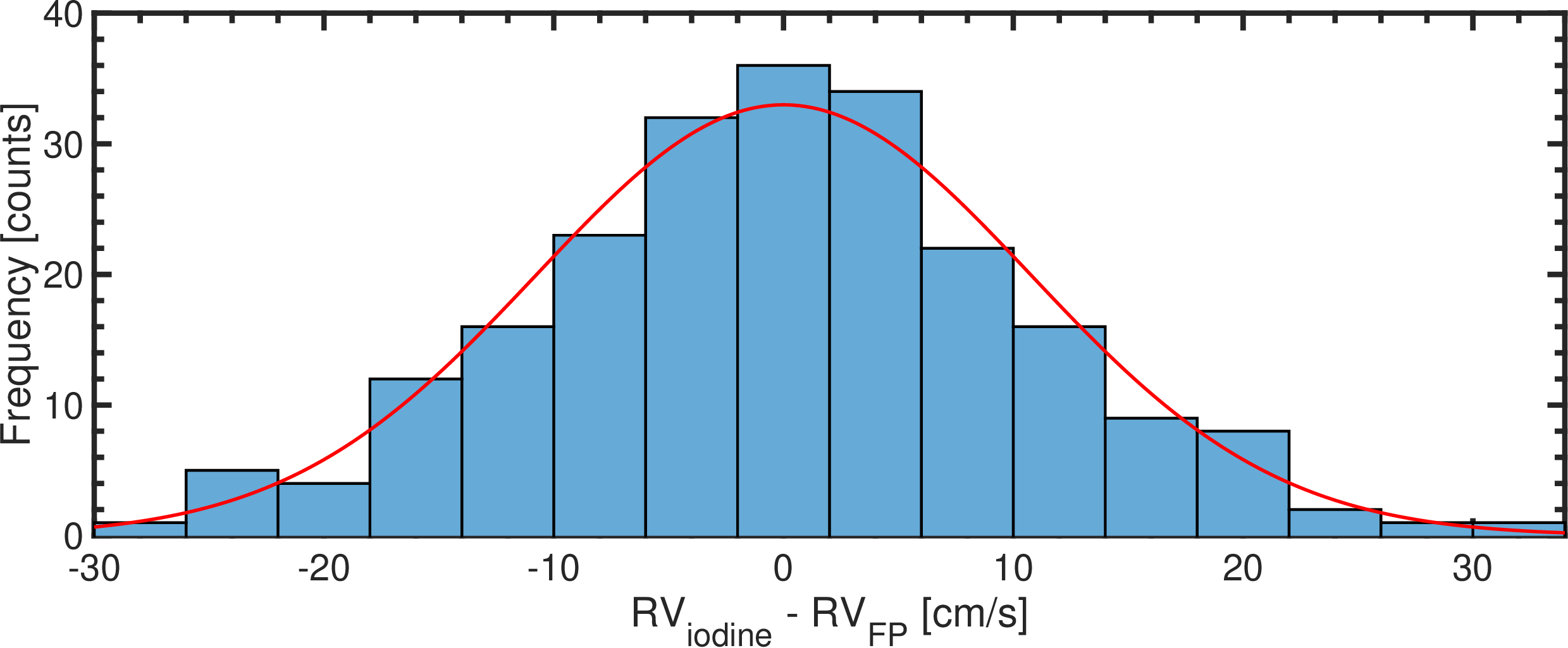}
 \caption{Histogram of the difference between the measured radial velocities of the FP and the iodine cell. The fitted normal distribution has a $\sigma$-value of \SI{10.7}{\centi\meter\per\second}.}\label{fig:rvdiffhist}
\end{figure}

We calculate the difference between the RV shift of the iodine cell and the FP (see Fig.\ \ref{fig:rvdiff}). Both light sources show good agreement. The overall RMS in the RV difference of \SI{10.7}{\centi\meter\per\second} corresponds well with the average uncertainty of the RV difference of \SI{10.2}{\centi\meter\per\second}. A histogram of the RV difference shows good agreement with a normal distribution (see Fig.\ \ref{fig:rvdiffhist}), indicating that we are dominated by random measurement noise, mainly caused by the higher RV uncertainty of the iodine cell. 

No significant drift between both light sources is apparent. The linear fit to the RV differences has a slope of \SI{0.9 \pm 1.3}{\centi\meter\per\second\per\day}, consistent with zero. Since the RVs of both light sources in the different input channels of the FTS show such good agreement, we also conclude that if there is any drift between both input channels it is negligible. The (non significant) value of the drift is in the same order of magnitude as the bulk shift measured in the FPs of HPF and ESPRESSO,\cite{Terrien2021,Schmidt2022} however we need long term measurements to confirm whether we also see such a drift in our FP.

%-no significant gls periods detectable, particularly no peak at a 1 day period.
To rule out any influence of the laboratory or vacuum vessel temperatures on the measured RVs we calculate the correlation between measured RVs and temperatures. For this we calculate the mean temperature for each sensor during each measurement. For the RV-difference we do not find any significant correlation with the measured temperatures. For the individual RVs (the values are similar for both the FP RVs and iodine RVs as well as their average) we find significant correlations with the room temperature and vacuum vessel temperatures. Since it is unlikely that both the FP and the iodine cell RVs are affected at the same level by the changing room or FP vacuum vessel temperature, the best explanation is that the FTS itself - most likely the HeNe reference laser frequency - is affected by the changing room temperature rather than either of the light sources. While there may be small non-white noise variations in the RV difference data, e.g., at 65 hours, calculating a GLS-periodogram did not yield any significant periods.
%The room temperature also seems to have a negative correlation with the temperature of the vacuum vessel. This may be caused by the backflowing thermooil being affected by the room temperature. We therefore changed to a process temperature control, i.e. using a temperature sensor in the vacuum vessel (the one located at the tank wall) for temperature control, rather than the internal sensor at the thermostat. 

\section{Discussion}% \& Outlook}
\label{sec:conc}
We show that that the FTS is intrinsically stable to a level of $\pm\,$\SI{1}{\meter\per\second} on timescales of days in periods of low room temperature variability ($<\!$\SI{1}{\kelvin}). Furthermore, we can calibrate the RV drift of the FTS to a precision of \SI{10.7}{\centi\meter\per\second}. The average intrinsic RV precision of a single measurement of the FP is on a level of \SI{1.0}{\centi\meter\per\second}. 
Currently, the RV difference uncertainty between both channels is dominated by the uncertainty in the RVs of the iodine cell (\SI{10.2}{\centi\meter\per\second}). This may be improved further by using a light source with more flux in the relevant wavelength regions, e.g., an LED.
On a timescale of 90 hours no significant drift between the iodine cell and the FP, and therefore also between the two wavelength channels of the FTS, is apparent. The RVs measured from both light sources show excellent agreement, with the difference in RV drift between both light sources following a normal distribution insignificantly broader than the expected uncertainty (\SI{10.7}{\centi\meter\per\second} and \SI{10.2}{\centi\meter\per\second}, respectively). Therefore, we conclude that only a negligible systematic variation between both channels remains on timescales of days.

The spectrum of molecular iodine is not only a relative reference, but an absolute one. A theoretical model of the spectrum of iodine with a 2$\sigma$ uncertainty better than \SI{3}{\mega\hertz} is available in the IodineSpec software package. \cite{Knoeckel2004} The model was created by fitting molecular potentials using experimentally measured transition frequencies and therefore depends on the accuracy of the measured transition frequencies, which had an uncertainty of down to \SI{3}{\mega\hertz}. This model has been experimentally verified for multiple transitions each at different wavelengths (\SIlist[list-units = single]{534;535;548;560;647;671}{\nano\meter}) using LFCs. \cite{Zhang2009,Hsiao2013,Huang2013,Shie2013,Huang2018,Cheng2019} The deviations to the model range from \SIrange{-3.6}{2.2}{\mega\hertz} (corresponding to better than \SI{2.5}{\meter\per\second}) with measurement uncertainties typically in the range of \SIrange{10}{25}{\kilo\hertz} (corresponding to better than \SI{2}{\centi\meter\per\second}). While this shows that the model does not (yet) exhibit accuracy on the sub \SI{}{\meter\per\second} level, the spectrum of iodine itself provides an accurate standard as illustrated by its recommended use as a practical realization of the meter by CIPM. \cite{riehle2018cipm}
While iodine transition frequencies shift with pressure, the constraints this puts on the cell temperature are easily manageable. Pressure shifts of iodine lines are found to be in the range of \SIrange{5}{9}{\kilo\hertz\per\pascal}.  \cite{Zhang2009,Hsiao2013,Huang2013,Shie2013,Huang2018,Cheng2019} With the vapor pressure temperature dependence of iodine being \SI{14.9}{\pascal\per\kelvin} at about \SI{45}{\celsius}  \cite{Berkenblit1966} (and increasing for higher temperatures) the pressure shift as a function of cell temperature can be estimated to be less than \SI{10}{\centi\meter\per\second\kelvin}, consistent with FTS measurements. \cite{Perdelwitz2018} This means that by maintaining a temperature stability of better than \SI{1}{\kelvin}, a \SI{10}{\centi\meter\per\second} accuracy of the iodine spectrum can be achieved. 
We therefore posit, that iodine can be used as an accurate frequency calibrator on a \SI{10}{\centi\meter\per\second} level. Although the number of transitions with sufficiently low uncertainty or models of the iodine spectra may not yet be sufficient to transfer this accuracy onto the SI standard, the actual spectrum of iodine can be a standard. With measurements at all relevant frequencies enabled by LFCs, an atlas at this level of accuracy can be created in the future. The measurement precision we achieved in this work exhibits the high spectral information content of molecular iodine. Therefore, we argue that RV measurements at a level of \SI{10}{\centi\meter\per\second} precision and accuracy are feasible with an FTS calibrated using an iodine cell.

Our measurements show no significant relative drifts between both calibration light sources and therefore the science and the calibration channel. Thus, we can use the FP to transfer the absolute calibration with the iodine cell from preceding or subsequent measurements onto solar measurements with the FP in parallel. We can not exclude an absolute offset between both channels. But since both channels drift in unison - at least on timescales of days - we only need an absolute calibration for the science channel, while using the calibration channel only to measure relative drifts.
%In our setup, the accuracy of the iodine spectrum can be transferred to solar measurements in the science channel via parallel FP measurements in the calibration channel. An absolute offset between the science and the calibration channel would have no influence on the accuracy of the solar measurements since the solar measurements are performed in the same channel as the iodine. The FP is only required to track the drift of the FTS and we showed that there is no relevant systematic drift between both channels.
Since the FP will exhibit an overall drift and a chromatic RV drift \cite{Terrien2021,Schmidt2022,Kreider2022} over time it has to be calibrated with the iodine cell at regular intervals. We expect weekly combined iodine cell and FP measurements will suffice to obtain solar spectra with an RV accuracy of \SI{10}{\centi\meter\per\second}.

%In the future, the iodine cell can also be used for an absolute wavelength calibration from theoretical spectra.\cite{Knoeckel2004} [sollen wir auch noch Takeda zitieren?] The FP may then be used for simultaneous calibration during measurements and can always be referenced to the iodine as required. This will yield not only high precision RV-measurements but also a high wavelength accuracy. Furthermore, we plan to monitor longterm changes in the FPs overall FSR and d-curve.

%We have shown a sufficient temperature stability on a timescale of one month. We are confident that the process temperature control (i.e. controlling the temperature with a sensor inside the vacuum vessel) will mitigate possible effects of seasonal temperature changes, leading to a constant FP temperature on timescales of years. If a better temperature stability of the FP is required in the future, the temperature cross correlation data suggests, that a better thermal contact between optical bench and FP-housing should improve temperature stability of the FP even further. A second or third heatshield layer would also be beneficial to reduce the impact of the non-stabilized lid.

\subsection* {Data availability} 
All data used in this work made available under https://doi.org/10.25625/PONFBL. FTS Spectra can be converted to other file formats using opus2py (https://gitlab.gwdg.de/karl.wessel/opus2py). 

\subsection* {Acknowledgments}  
MD thanks M. Zechmeister for the kind help regarding SERVAL.
The authors thank BMBF for funding (project 05A17MG3).

% References
\bibliography{lfc_fts.bib} % bibliography data in report.bib
\bibliographystyle{spiejour} % makes bibtex use spiebib.bst
%\bibliographystyle{spiebib} % makes bibtex use spiebib.bst
%\end{spacing}
\end{document}